\documentclass[pdflatex,sn-mathphys-num]{sn-jnl}

\usepackage{graphicx}%
\usepackage{multirow}%
\usepackage{amsmath,amssymb,amsfonts}%
\usepackage{amsthm}%
\usepackage{mathrsfs}%
\usepackage[title]{appendix}%
\usepackage{xcolor}%
\usepackage{textcomp}%
\usepackage{manyfoot}%
\usepackage{booktabs}%
\usepackage{algorithm}%
\usepackage{algorithmicx}%
\usepackage{algpseudocode}%
\usepackage{listings}%

\begin{document}

\title[Article Title]{Localization Properties of a Disordered Helical Chain}

\author[1]{\fnm{Taylan} \sur{Yildiz}}

\author*[1]{\fnm{B.} \sur{Tanatar}}\email{tanatar@fen.bilkent.edu.tr}

\affil[1]{\orgdiv{Department of Physics}, \orgname{Bilkent University}, \city{Ankara}, 06800 T\"urkiye}

\abstract{We study the localization properties of the quasiperiodic one-dimensional helical chain with two tunneling paths: nearest-neighbor and a long-range hop that connects sites of consecutive helical turns. Using exact diagonalization, we quantify localization employing the inverse participation ratio (IPR) and the normalized participation ratio (NPR), and combine them into a single measure to create a phase map. The resulting diagrams reveal three regimes: a completely extended phase, a completely localized phase, and a mixed domain where localized and extended states coexist. In the diagrams, we investigate the behaviors of tightly and loosely wound helices and examine a special case where the number of sites per turn is a Fibonacci number. For moderate numbers of sites per helical turn, the mixed region is broad and also shifts with the long-range coupling. When the turn size is a Fibonacci number, the phase boundary becomes nearly horizontal and the mixed region fades out, effectively recovering the standard Aubry-Andr\'e model behavior. }


\maketitle

\section{Introduction}\label{sec1}
The localization phenomenon has been the subject of great interest in condensed matter physics for many years \cite{1,2}. Anderson's seminal paper \cite{3} initiated the field by showing that a random potential localizes the single-particle wavefunction.  While Anderson Localization (AL), uncorrelated disorder, localizes all single-particle states, the quasiperiodically modulated potentials, such as the celebrated Aubry-Andr\'e (AA) model \cite{4,5}, can realize a tunable metal-insulator transition without introducing true randomness.

In one dimension with uncorrelated (random) disorder, the Anderson model predicts exponential localization of all single-particle eigenstates for arbitrarily weak disorder. By contrast, the AA model employs a quasiperiodic onsite modulation $V_n=\Delta\cos(2\pi\beta n+\phi)$ and possesses an exact self-duality at $\Delta/J=2$ (with $J$ the nearest-neighbor hopping), all eigenstates transition collectively from extended ($\Delta/J<2$) to localized ($\Delta/J>2$). 

Departures from exact AA self–duality with deformations of the quasiperiodic potential or longer range hoppings render the localization threshold energy dependent, producing mobility edges where localized and extended states coexist within the same spectrum \cite{11,18,19}. In our helical setting, the appearance of the mobility edge with $N$ and $J_N$ underlies the mixed-phase regions, highlighted in our diagrams.

Localization transitions related to the Anderson and AA scenarios have been demonstrated in ultracold atoms using bichromatic optical lattices, where the quasiperiodic amplitude $\Delta$ and hopping $J$ are independently tunable, and in a photonic waveguide arrays, where light propagation images the spatial profile of eigenmodes in real space \cite{6,7,8,9,20}. The present helical long-range coupling can be engineered within these settings, in photonics, by engineering
waveguides or ring resonators on a helical trajectory so that  coupling across a turn provides the $N$th-neighbor link or
in cold atoms, by using Raman-assisted tunneling to connect site $n$ to $n\!+\!N$ along a synthetic or real lattice direction.

Beyond the basic transition, the AA framework builds a broad literature on mobility edges and generalized couplings, including models with exact mobility edges, long-range hopping variants, non-Hermitian extensions, and also reentrant phenomena  \cite{10,12,13,135,136,21,22,23}. 

In this work, we extend the AA model to a one-dimensional helical chain by applying a quasiperiodic potential to a tight-binding representation of a helix. The geometry of the helix allows us to represent the structure by a long-range hopping, which we call the $N$th hopping, $N$ depending on the helical winding \cite{14,15,16}. A helix supplies a minimal, deterministic way to go beyond the AA model. It preserves 1D connectivity and the same quasiperiodic onsite modulation, but adds a single, tunable long–range channel (the $N$th hop) that breaks AA self–duality in a controlled manner. Unlike most generalizations that alter the onsite potential or introduce many long–range terms at once, the helical construction isolates one geometric parameter ($N$) and thereby cleanly tests the AA transition to structured long–range tunneling. Because $N$ competes with the incommensurate frequency $\beta$, the model also exposes commensurability effects that are largely missing from existing studies and that naturally promote energy–dependent localization (mobility edges). We solve the model for a spinless fermion by exact diagonalization, employing the use of IPR and NPR \cite{17} to quantify localization behavior. We identify the extended, localized, and mixed regimes by constructing a phase diagram with respect to quasiperiodic potential strength. Localization in helical lattice geometries has been explored in previous works \cite{24,25}, which further motivates the present framework. 

\section{Model and Approach}\label{sec2}

\begin{figure}[h!]
    \centering
    \includegraphics[width=119mm]{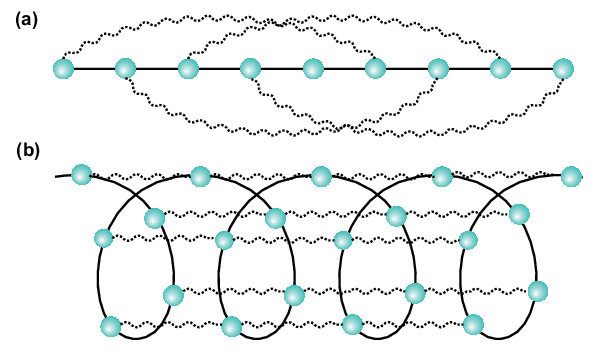}
    \caption{Schematic representation of helical tight-binding model. (a) One-dimensional chain representation of the long-range hopping model. Circles illustrate lattice sites, solid lines represent nearest neighbor hopping, and dotted arches show long-range hopping that connects a site to the one located in the next helical winding ($N$th neighbor, here $N=5$). (b) Geometric picture of the same model in an unwrapped helix. Sites are arranged with $5$ lattice points per turn.}
    \label{fig1}
\end{figure}

We consider a 1D lattice with nearest-neighbor and $ N$th-neighbor hopping, subject to a quasiperiodic potential, described by the Hamiltonian
\begin{align}
     H=-\sum_{n=1}^{L-1}Jc_{n}^\dagger c_{n+1}+{\rm h.c.}-\sum_{n=1}^{L-N}J_Nc_{n}^\dagger c_{n+N}+{\rm h.c.}+\sum_{n=1}^{L} \Delta\cos(2\pi\beta n)c_{n}^\dagger c_{n}
\end{align}
where $c_n^{\dagger}$ ($c_n$) creates (annihilates) a fermion on site $n$. Total length of the chain is $L$ while $J$ and $J_N$ are the nearest-neighbor and $N$th-neighbor hopping strengths, respectively. The second term in the Hamiltonian is the onsite quasiperiodic potential. Here, $\Delta$ is the potential strength and $\beta$ is the irrational number to ensure incommensurability along the chain. The $N$th hopping term in the Hamiltonian gives rise to the helical geometry, as can be seen in Fig.\ref{fig1}. In a helical chain with $N$ lattice points in each winding, the $N$th hopping represents the tunneling between the successive helical windings. 

We use IPR and NPR to examine localization properties. IPR and NPR of the $i$th eigenstate are defined as follows.
\begin{align}
    {\rm IPR}_i=\sum_{n=1}^L|\psi_i^{(n)}|^4, \quad \quad {\rm NPR}_i=\bigg(L\sum_{n=1}^L|\psi_i^{(n)}|^4\bigg)^{-1}
\end{align}
In our phase diagrams, we use average IPR and NPR in a composite measure $\eta$ to capture the overall localization behavior of the system \cite{16}. 
\begin{align}
    \eta=\log_{10}(\langle {\rm IPR}\rangle\times\langle {\rm NPR}\rangle)
\end{align}
When the spectrum is (almost) entirely extended or entirely localized, we have \(\eta < -\log_{10}L\). 
When localized and extended states coexist (mixed regime), \(-2 \lesssim \eta \lesssim -1\). Since, in the localized regime \(\langle \mathrm{IPR}\rangle = O(1)\) and \(\langle \mathrm{NPR}\rangle \sim 1/L\); in the extended regime \(\langle \mathrm{NPR}\rangle = O(1)\) and \(\langle \mathrm{IPR}\rangle \sim 1/L\); and in the mixed regime both averages remain \(O(1)\). 
Because \(\eta\) is symmetric under \(\mathrm{IPR}\!\leftrightarrow\!\mathrm{NPR}\), it cannot distinguish a purely extended phase from a purely localized one. 
To resolve this ambiguity, we examine the individual indicators \(\mathrm{IPR}\) and \(\mathrm{NPR}\).

\section{Results and Discussion}\label{sec3}
In our numerical calculations, we scale energies by $J$, use a system size $L=2584$, and cover three different helical turns $N=40$ (tightly wound), $N=200$ (loosely wound), and $N=377$ (Fibonacci wound). In compact helices, $N$th-neighbor can have short spatial separation relative to the nearest-neighbor, so the $N$th-neighbor hopping \(J_N\) can be of the same order as the nearest-neighbor hopping \(J\). Thus, for $J_N$, we span values from $0$ to $2J$.
\begin{figure}[h!]
    \centering
    \includegraphics[width=119mm]{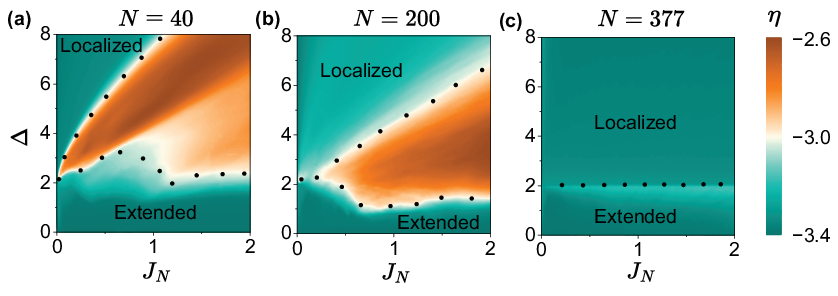}
    \caption{Phase diagrams of the exponent $\eta$ in $\Delta$ and $J_N$ plane for different number of helical turns. (a) $N=40$ (b) $N=200$ (c) $N=377$. Green color denotes a completely localized or extended regime, whereas orange color indicates a mixed regime where localized and extended regimes coexist. The black dots mark the estimated phase boundaries.}
    \label{fig2}
\end{figure}

In Fig.\ref{fig2}, we present phase diagrams revealing the $N$-dependence. For $N=40$ Fig.\ref{fig2}(a) and $N=200$ Fig.\ref{fig2}(b), we see that completely localized and extended regions (green) are clearly separated by a mixed region (orange). For $N=40$, a broad mixed region appears after $\Delta\approx2$, signaling a mobility-edge with coexistence of localized and extended states. As the helix loosens $N=200$, the boundaries shift to smaller $\Delta$ values, and it becomes easier to completely localize the system. However, the mixed region still exists between the localized and extended regimes. A different pattern emerges when we let $N$ be a Fibonacci number. In Fig.\ref{fig2}(c), we see that when $N=377$, the mixed domain is essentially suppressed and the phase boundary becomes nearly horizontal around $\Delta=2$, weakly dependent on $J_N$. This trend arises from an arithmetic relation between the helical turn $N$ and the incommensurate frequency $\beta$. The $N$th-neighbor hopping connects sites whose onsite potential differs by 
\begin{align}
    \delta \Phi=\Delta\big\{\cos[2\pi\beta n]-\cos[2\pi\beta(n+N)]\big\} =2\Delta\sin(\pi\beta N)\sin(2\pi\beta n+\pi\beta N)
\end{align}
It is known that $\beta\approx F_{n-1}/F_n$ where $F_n$ is the $n$th Fibonacci number. When $N=F_n$, we find that 
\begin{align}
    \delta\Phi\approx\sin\bigg(\pi\frac{F_{n-1}}{F_n}F_n\bigg)=\sin(\pi F_{n-1})=0\;.
\end{align}
Hence, the $N$th hop couples the sites with nearly identical onsite energies and does not contribute to the localization; consequently, the phase boundary becomes horizontal, independent of $J_N$, near $\Delta\approx2$, mimicking the original AA model. 

\begin{figure}[h]
    \centering
    \includegraphics[width=119mm]{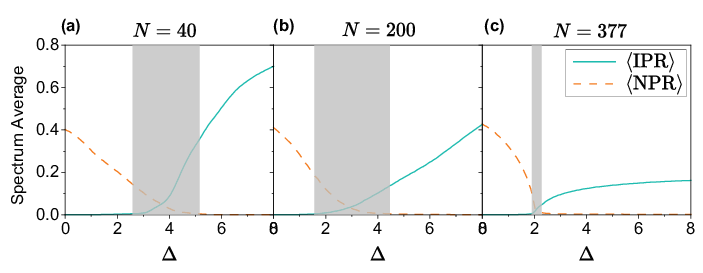}
    \caption{($J_N=0.5$): Spectrum averaged IPR and NPR versus quasi-periodic potential strength $\Delta$ for helical turns (a) $N=40$, (b) $N=200$, (c) $N=377$. Solid curves shows $\langle\rm IPR\rangle$ whereas dashed curves shows $\langle\rm NPR\rangle$. Gray vertical bands identify the mixed regions.}
    \label{fig3} 
\end{figure}
Fig.\ref{fig3} represents the spectrum averaged localization measures $\langle\rm IPR\rangle$ and $\langle\rm NPR\rangle$ with respect to $\Delta$ at fixed $J_N=0.5$. For small $\Delta$, each panel shows an extended spectrum with finite $\langle\rm NPR\rangle$ and nearly vanishing $\langle\rm IPR\rangle$; upon entering the gray window, each of the two measures assumes intermediate values, showing the coexistence of localized and extended states with mobility edges, which is in agreement with Fig.\ref{fig2}. For larger $\Delta$, the spectrum becomes localized as $\langle\rm NPR\rangle\rightarrow0$. The evolution with $N$ mirrors the phase diagrams. The mixed region is broad when $N=40$, it narrows with $N=200$, and almost vanishes with $N=377$, a Fibonacci number. 

\section{Conclusion}\label{sec4}
In this paper, we studied a helical (longer-range) extension of the well-known Aubry-Andr\'e model with $N$th hopping. Covering different helical windings $N$, we identified the phase maps of the model in the $J_N$ and $\Delta$ plane. The diagrams show three regions: a completely extended regime at small $\Delta$, a completely localized regime at high $\Delta$ and at moderate $\Delta$, and a mixed regime (mobility-edge behavior) where localized and extended states coexist. This coexistence is governed by the arithmetic relation between helical winding $N$ and the incommensurate frequency $\beta$. When $N$ is a Fibonacci number, the $N$th hop couples sites with nearly aligned onsite potentials, leaving the effective strength essentially unchanged. Therefore, the phase boundary becomes nearly horizontal near $\Delta\simeq 2$, weakly dependent on $J_N$. Overall, our results identify an arithmetic control knob for localization in helical lattices: while primarily the quasiperiodic potential strength controls the phases commensurability between $N$ and $\beta$ tunes how long-range hopping renormalizes the effective disorder. This suggests several directions for future work, including finite-size scaling, transport at controlled filling, many-body effects, and realizations in photonic or cold-atom platforms.

\section*{Statements and Declarations}
\subsection*{Declarations} The authors have no competing interests to declare relevant to this article's content.
\subsection*{Funding}
This work was supported in part by the Turkish Academy of Sciences (TUBA).
\subsection*{Data Availability}
Data sets generated during the current study are available from the corresponding author on reasonable request.

\bibliography{sn-bibliography}

@article{1,
  title={Localization: theory and experiment},
  author={Kramer, Bernhard and MacKinnon, Angus},
  journal={Rep. Prog. Phys.},
  volume={56},
  number={12},
  pages={1469},
  year={1993},
  publisher={IOP Publishing},
doi={10.1088/0034-4885/56/12/001}
}

@article{2,
  title={Anderson transitions},
  author={Evers, Ferdinand and Mirlin, Alexander D},
  journal={Rev. Mod. Phys.},
  volume={80},
  number={4},
  pages={1355--1417},
  year={2008},
  publisher={APS},
  doi = {10.1103/RevModPhys.80.1355}
}

@article{3,
	title = {Absence of {D}iffusion in {C}ertain {R}andom {L}attices},
	author = {Anderson, P. W.},
	journal = {Phys. Rev.},
	volume = {109},
	number = {5},
	pages = {1492--1505},
	year = {1958},
	doi = {10.1103/PhysRev.109.1492},
}

@article{4,
	title={Analyticity breaking and {A}nderson localization in incommensurate lattices},
	author={Aubry, Serge and Andr{\'e}, Gilles},
	journal={Ann. Israel Phys. Soc},
	volume={3},
    year={1980},
	number={133},
	pages={18}
}

@article{5,
	doi = {10.1088/0370-1298/68/10/304},
	year = {1955},
	volume = {68},
	number = {10},
	pages = {874},
	author = {P G Harper},
	title = {Single {B}and {M}otion of {C}onduction {E}lectrons in a {U}niform {M}agnetic {F}ield},
	journal = {Proc. Phys. Soc. A},
}

@article{6,
	title = {Anderson localization of a non-interacting {Bose}–{Einstein} condensate},
	volume = {453},
	doi = {10.1038/nature07071},
	number = {7197},
	journal = {Nature},
	author = {Roati, Giacomo and D’Errico, Chiara and Fallani, Leonardo and others},
	year = {2008},
	pages = {895--898},
}

@article{7,
	author = {Michael Schreiber and Hodgman, Sean S and Bordia, Pranjal and others},
	title = {Observation of many-body localization of interacting fermions in a quasirandom optical lattice},
	journal = {Science},
	volume = {349},
	number = {6250},
	pages = {842-845},
	year = {2015},
	doi = {10.1126/science.aaa7432},
}

@article{8,
	title = {Direct observation of {Anderson} localization of matter waves in a controlled disorder},
	volume = {453},
	doi = {10.1038/nature07000},
	number = {7197},
	journal = {Nature},
	author = {Billy, Juliette and Josse, Vincent and Zuo, Zhanchun and others},
	year = {2008},
	pages = {891--894},
}

@article{9,
  title={Observation of a localization transition in quasiperiodic photonic lattices},
  author={Lahini, Yoav and Pugatch, Rami and Pozzi, Francesca and others},
  journal={Phys. Rev. Lett.},
  volume={103},
  number={1},
  pages={013901},
  year={2009},
doi = {10.1103/PhysRevLett.103.013901}
}

@article{10,
	title = {Mobility edge and intermediate phase in one-dimensional incommensurate lattice potentials},
	author = {Li, Xiao and Das Sarma, S.},
	journal = {Phys. Rev. B},
	volume = {101},
	number = {6},
	pages = {064203},
	year = {2020},
	doi = {10.1103/PhysRevB.101.064203},
}

@article{11,
   title={Predicted {M}obility {E}dges in {One-Dimensional} {I}ncommensurate {Optical Lattices: An Exactly Solvable Model of Anderson Localization}},
   volume={104},
   DOI={10.1103/physrevlett.104.070601},
   number={7},
   journal={Phys. Rev. Lett.},
   author={Biddle, J. and Das Sarma, S.},
   year={2010}
}

@article{12,
  title={One-dimensional quasicrystals with power-law hopping},
  author={Deng, X and Ray, S and Sinha, S and Shlyapnikov, GV and Santos, L},
  journal={Phys. Rev. Lett.},
  volume={123},
  number={2},
  pages={025301},
  year={2019},
  doi = {10.1103/PhysRevLett.123.025301}
}

@article{13,
  title={Localization in one-dimensional lattices with non-nearest-neighbor hopping: {Generalized Anderson and Aubry-Andr{\'e}} models},
  author={Biddle, J and Priour Jr, DJ and Wang, B and Das Sarma, S},
  journal={Phys. Rev. B},
  volume={83},
  number={7},
  pages={075105},
  year={2011},
  doi ={10.1103/PhysRevB.83.075105}
}

@article{14,
  title={Bloch dynamics in lattices with long-range hopping},
  author={Stockhofe, J and Schmelcher, P},
  journal={Phys. Rev. A},
  volume={91},
  number={2},
  pages={023606},
  year={2015},
  doi={10.1103/PhysRevA.91.023606}
}

@article{15,
  title={Bethe ansatz study of 1+1 dimensional {H}ubbard model},
  author={Xiong, Shi-Jie},
  journal={Z. Phys. B},
  volume={89},
  number={1},
  pages={29--34},
  year={1992},
  doi={10.1007/BF01320825}
}

@article{16,
  title={Possible first order phase transition in the one-dimensional helical {H}ubbard model},
  author={Wang, Wenzheng and Xiong, Shi-jie},
  journal={Phys. Lett. A},
  volume={156},
  number={7-8},
  pages={415--418},
  year={1991},
  doi={10.1016/0375-9601(91)90719-O}
}

@article{17,
  author  = {Wegner, F.},
  title   = {Inverse participation ratio in $2 + \epsilon$ dimensions},
  journal = {Z. Phys. B},
  volume  = {36},
  number  = {3},
  pages   = {209--214},
  year    = {1980},
  doi     = {10.1007/BF01325284},
}

@article{135,
  author  = {Sajid, M and Khan, N A and Shah, M},
  title   = {Topological pumping in an inhomogeneous {A}ubry–{A}ndr{\'e} model},
  journal = {Chin. J. Phys.},
  volume  = {92},
  pages   = {311-320},
  year    = {2024},
  doi     = {10.1016/j.cjph.2024.09.028},
}

@article{136,
  author  = {Souvik, R and Maiti, S R and Laroze, D},
  title   = {Anomalous persistent current in a 1D dimerized ring with aperiodic site potential: {N}on-interacting and interacting cases},
  journal = {Chin. J. Phys.},
  volume  = {96},
  pages   = {542-558},
  year    = {2025},
  doi     = {10.1016/j.cjph.2025.05.024},
}

@article{18,
  title={Nearest neighbor tight binding models with an exact mobility edge in one dimension},
  author={Ganeshan, Sriram and Pixley, JH and Das Sarma, S},
  journal={Phys. Rev. Lett.},
  volume={114},
  pages={146601},
  year={2015},
  doi={10.1103/PhysRevLett.114.146601}
}

@article{19,
  title={Energy-dependent dynamical quantum phase transitions in quasicrystals},
  author={Ye, Shihao and Zhou, Ziheng and Khan, Niaz Ali and Xianlong, Gao},
  journal={Phys. Rev. A},
  volume={109},
  pages={043319},
  year={2024},
  doi={10.1103/PhysRevA.109.043319}
}

@article{20,
  title={Transport and Anderson localization in disordered two-dimensional photonic lattices},
  author={Schwartz, Tal and Bartal, Guy and Fishman, Shmuel and Segev, Mordechai},
  journal={Nature},
  volume={446},
  pages={52--55},
  year={2007},
  doi={10.1038/nature05623}
}

@article{21,
  title={Coexistence of reentrant localization and dynamical delocalization in a one-dimensional non-Hermitian quasiperiodic lattice},
  author={Wang, Haoyu and Zheng, Xiaohong and Xiao, Liantuan and Jia, Suotang and Chen, Jun and Zhang, Lei},
  journal={Phys. Rev. B},
  volume={112},
  pages={054202},
  year={2025},
  doi={10.1103/bd1n-dclq}
}

@article{22,
  title={Complete delocalization and reentrant topological transition in a non-Hermitian quasiperiodic lattice},
  author={Padhan, Ashirbad and Padhi, Soumya Ranjan and Mishra, Tapan},
  journal={Phys. Rev. B},
  volume={109},
  pages={L020203},
  year={2024},
  doi={10.1103/PhysRevB.109.L020203}
}

@article{23,
  title={Reentrant localization transition in a quasiperiodic chain},
  author={Roy, Shilpi and Mishra, Tapan and Tanatar, Bilal and Basu, Saurabh},
  journal={Phys. Rev. Lett.},
  volume={126},
  pages={106803},
  year={2021},
  doi={10.1103/PhysRevLett.126.106803}
}

@article{24,
  title={Phenomenon of multiple reentrant localization in a double-stranded helix with transverse electric field},
  author={Ganguly, Sudin and Sarkar, Suparna and Mondal, Kallol and Maiti, Santanu K},
  journal={Scientific Reports},
  volume={14},
  pages={3059},
  year={2024},
  doi={10.1038/s41598-024-52579-2}
}

@article{25,
  title={Critical analysis of multiple reentrant localization in an antiferromagnetic helix with transverse electric field: Hopping dimerization-free scenario},
  author={Ganguly, Sudin and Chattopadhyay, Sourav and Mondal, Kallol and Maiti, Santanu K},
  journal={SciPost Physics Core},
  volume={8},
  pages={012},
  year={2025},
  doi={10.21468/SciPostPhysCore.8.1.012}
}
\end{document}